\documentclass[prc,aps]{revtex4}
\usepackage{graphicx}

\begin{document}
\title{Search for the Production of Element 112 in the $^{48}$Ca + $^{238}$U
Reaction}
\author{W.~Loveland$^{1}$, K.E.~Gregorich$^{2}$, J.B.~Patin$^{2}$,
  D.~Peterson$^{1}$, C.~Rouki$^{3}$, P.M.~Zielinski$^{2}$ and K.~Aleklett$^{3}$}
\affiliation{$^{1}$Dept. of Chemistry, Oregon State University, Corvallis,OR 97331}
\affiliation{$^{2}$Nuclear Science Division, Lawrence Berkeley National\\
Laboratory,Berkeley, CA 94720}
\affiliation{$^{3}$Uppsala University, Uppsala, Sweden}
\date{\today}

\begin{abstract}
We have searched for the production of element 112 in the reaction of 231
MeV $^{48}$Ca with $^{238}$U. We have not observed any events with a ``one
event'' upper limit cross section of 1.6 pb for EVR-fission events and 1.8
pb for EVR-alpha events.
\end{abstract}


\maketitle

\section{Introduction}

\bigskip The heaviest elements are a laboratory to study nuclear structure
and nuclear dynamics under the influence of large Coulomb forces. The
results of heavy element research deal with fundamental issues in both
chemistry and physics. During the past six years, there have been
spectacular advances in this field, {\it i.e.}, the discovery of elements
110-112, the synthesis of elements 114 \cite{yuri1,yuri2} and element 116 
\cite{yuri3} by ``hot fusion'' reactions, the first chemical studies of
elements 104-108 and the spectroscopy of the transfermium nuclei.

\begin{figure}[tbp]
	\includegraphics[width=.8\textwidth]{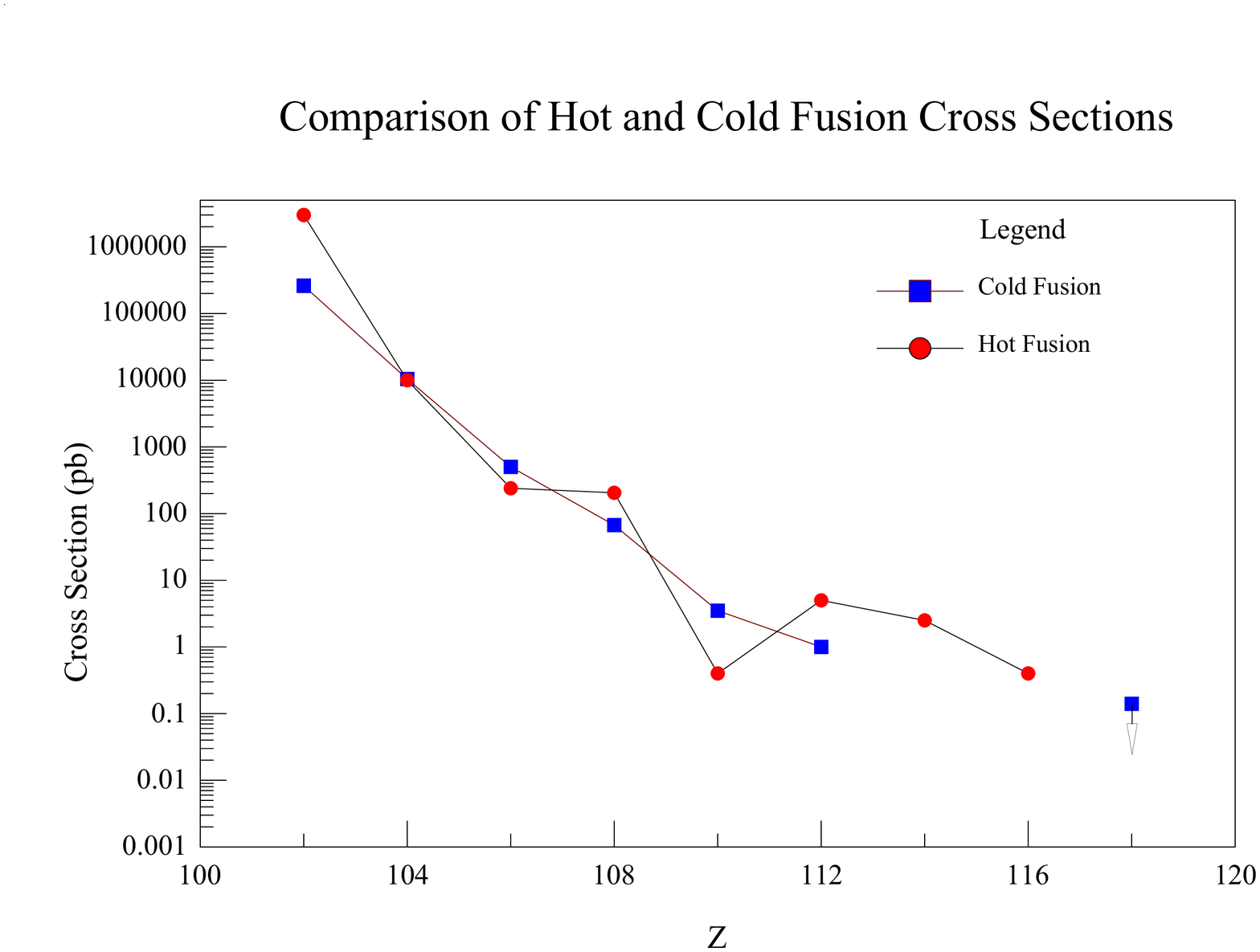}
	\caption{The predicted and observed cross sections for the synthesis of
	  heavy nuclei using ``hot" and ``cold" fusion reactions. The value shown
	  for element 118 is an upper limit.\label{fig1}}
\end{figure}

As an aside, we note the two different traditional paths to the heavy
elements: (a) ``cold fusion'', involving the reaction of massive projectiles
with Pb and Bi target nuclei, leading to low excitation energies in the
completely fused species (with resulting high survival probabilities) and
reduced fusion cross sections and (b) ``hot fusion'', the reaction of
lighter projectiles with actinide target nuclei, leading to larger fusion
cross sections but reduced survival probabilities (due to the higher
excitation energies of the completely fused species.) At present, it appears
that hot fusion reactions are the preferred path to synthesize new heavy
elements (Figure \ref{fig1}) although the large cross sections associated
with the production of elements 112-116 are poorly understood \cite{russ1}.
In any case, it is imperative to confirm these reported hot fusion cross
sections in laboratories not connected to the original work.

In 1999, a Dubna-GSI-RIKEN collaboration \cite{yuri4} reported the
successful synthesis of $^{283}$112 using the reaction 231 MeV $^{48}$Ca + 
$^{238}$U $\rightarrow ^{286}$112 $\rightarrow ^{283}$112 + 3n with the
observation of two events. The nuclide $^{283}$112 (t$_{1/2}$ = 
81$_{-32}^{+147}$ s) was reported to decay by spontaneous fission (SF) and was
produced with a cross section of 5.0$_{-3.2}^{+6.3}$ pb. The decay mode of 
$^{283}$112 is somewhat unexpected as all the other isotopes of element 112
(A=277,284 and 285) decay by alpha emission. The Dubna-GSI-RIKEN
collaboration searched for alpha decay in $^{283}$112 but did not see any
events.  Subsequently, in the reaction of $^{48}$Ca with $^{242}$Pu, two
events were found in which an evaporation residue (EVR) emitted an
alpha-particle, producing a daughter nucleus that decayed by SF. \cite
{yuriex}\ These latter SF decays were attributed to the decay of $^{283}$112
and if taken with the previous work, imply a half-life of$\sim$3 m for 
$^{283}$112. 

\begin{figure}[tbp]
	\includegraphics[width=.75\textwidth]{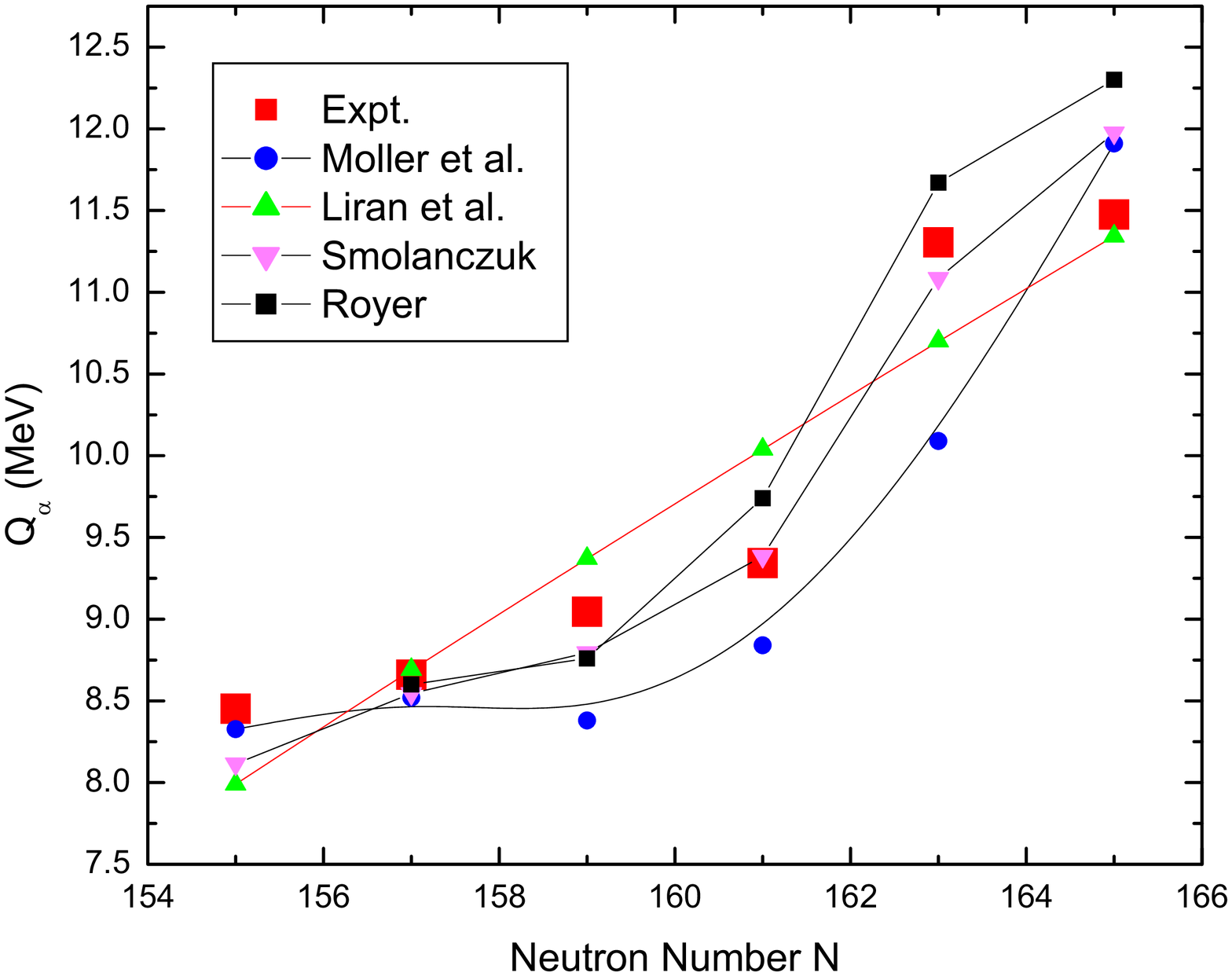}
	\caption{Predicted and measured Q$_{\protect\alpha}$ values for
	  the decay of $^{277}$112.\label{fig2}}
\end{figure}

\begin{figure}[tbp]
	\includegraphics[width=.75\textwidth]{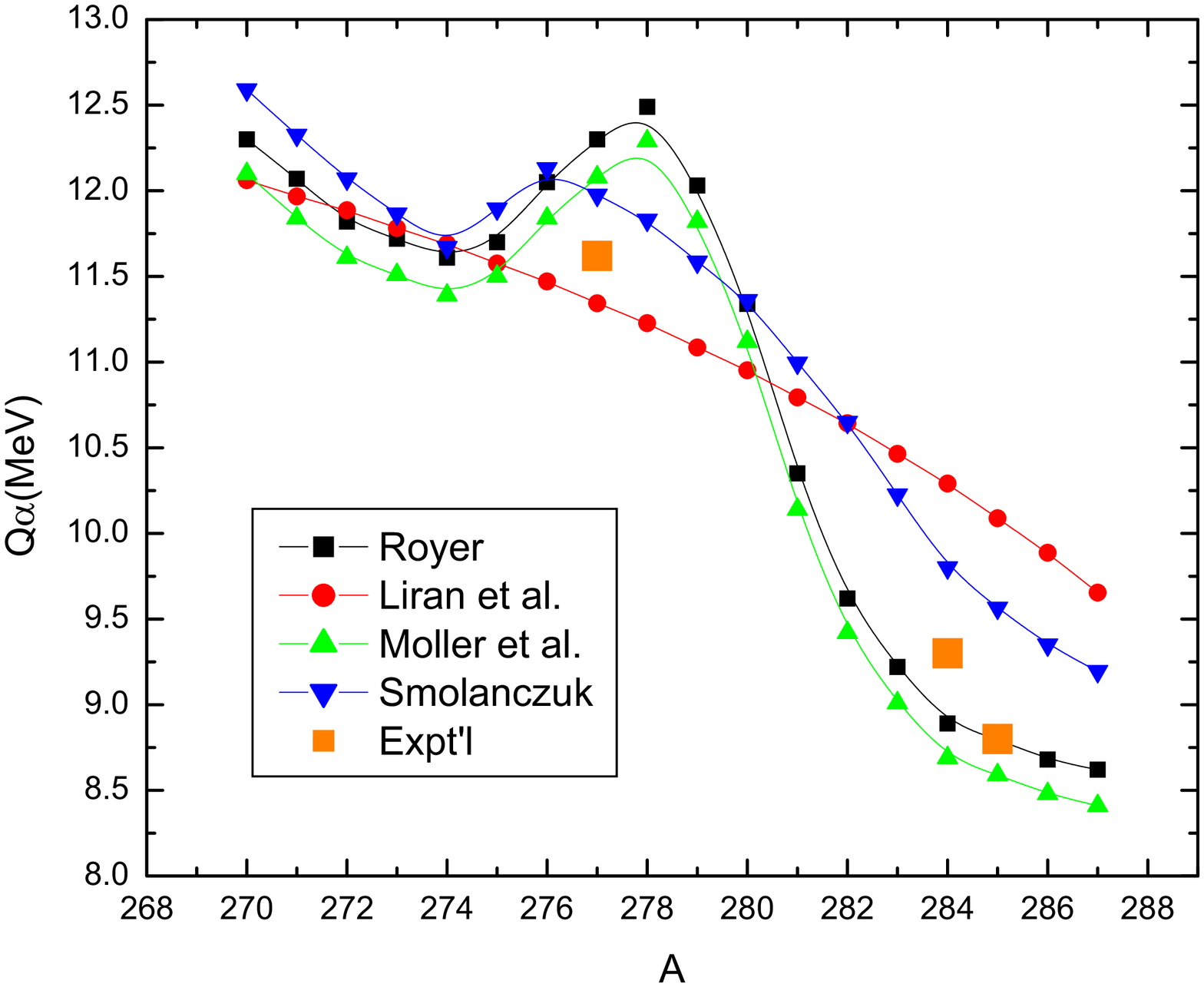}
	\caption{Predicted and measured Q$_{\protect\alpha}$ values for
	  the decay of various isotopes of element 112.\label{fig3}}
\end{figure}

In Figure \ref{fig2}, we show the predicted \cite{royer,liran,robert,peter}
and observed Q$_{\alpha}$ values for the well-characterized alpha-decay of 
$^{277}$112 and its daughters ($^{273}$110,$^{269}$Hs, $^{265}$Sg,
$^{261}$Rf and $^{257}$No).  The semi-empirical predictions of Liran {\it et
al.}\cite {liran} apparently do not include the nuclear structure effects
near the N=162 subshell.  The theoretical predictions of Smola\'{n}czuk
\cite{robert} seem to do the best job of predicting the observed values of
Q$_{\alpha}$ ($\chi_{M\ddot{o}ller}^{2} = 960$, $\chi_{Liran}^{2} = 400$, 
$\chi_{Smola\acute{n}czuk}^{2} = 160$, $\chi_{Royer}^{2} = 400$).  In Figure
\ref{fig3}, we show a similar plot of the predicted and observed values of 
Q$_{\alpha}$ for the $\alpha$-decay of various isotopes of element 112.
The predictions of Liran et al. deviate significantly from the observed
values with the predictions of Royer and M\"{o}ller et al. being similar. 
The theoretical predictions of M\"{o}ller {\it et al.} and Smola\'{n}czuk
are approximately equal in their ability to predict Q$_{\alpha}$ with a
slight preference being given to the predictions of M\"{o}ller {\it et al.} 
($\chi_{M\ddot{o}ller}^{2} = 240$, $\chi_{Liran}^{2} = 1080$, 
$\chi_{Smola\acute{n}czuk}^{2} = 400$, $\chi_{Royer}^{2} = 240$) Using these
comparisons of predicted and observed values of Q$_{\alpha}$ as a guide, we
favor the predictions of Smola\'{n}czuk as being the most reliable guide to
the expected decay properties of element 112.  However some caution must be
exercised as none of the predictions provide a statistically significant fit
to the data. In the only calculation \cite{robert} to address the
spontaneous fission and alpha decay of the isotopes of 112, alpha decay is
predicted to be the dominant mode of decay for all isotopes although the
differences in predicted half-lives are only an order of magnitude for the
nuclei of interest.

\begin{figure}[tbp]
	\includegraphics[width=.6\textwidth]{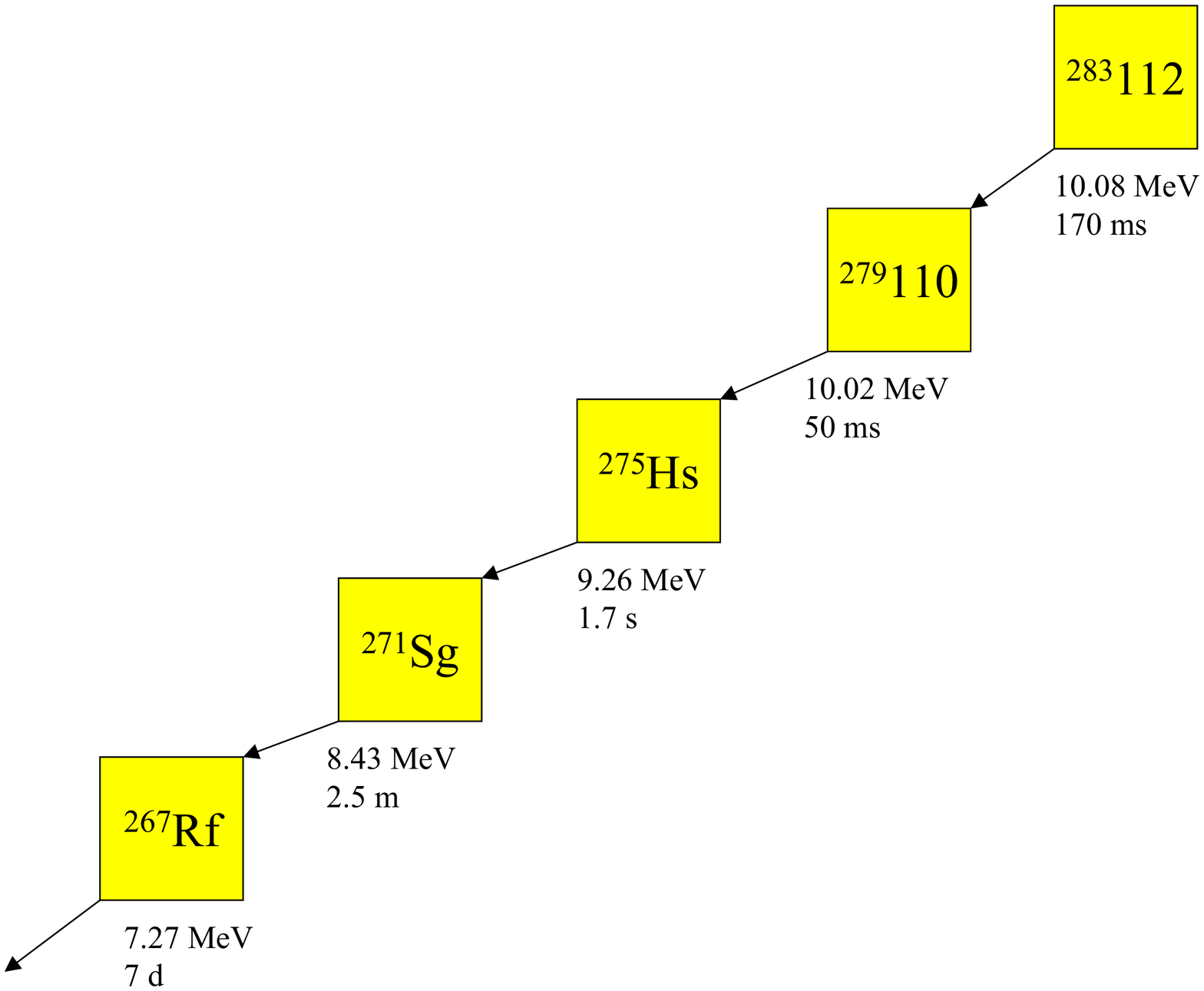}
	\caption{Predicted alpha-decay sequences for the decay of
	  $^{283}$112.\label{fig4}}
\end{figure}

We show in Figure \ref{fig4}, the expected alpha decay sequence for
$^{283}$112 based upon the predictions of Smola\'{n}czuk for the masses of
the heaviest elements and the Hatsukawa-Nakahara-Hoffman rules for the alpha
decay lifetimes of the heavy nuclei. \cite{hatsukawa}.  As indicated
earlier, in searching for these predicted alpha decay sequences, one must be
sensitive over a wide range of nuclear lifetimes.

The nucleus $^{283}$112 and its synthesis play an important role in our
understanding of the recent syntheses of elements 114 and 116 by hot fusion
reactions \cite{yuri1,yuri2,yuri3}. $^{283}$112 is directly populated in the
de-excitation of $^{287}$114 synthesized using the $^{48}$Ca + $^{242}$Pu
reaction\cite{yuriex}. The long half-life is typical of elements 112 and 114
nuclei produced in the synthesis of elements 114 and 116 
\cite{yuri1,yuri2,yuri3}. The relatively large reported production cross
section, 5 pb, is typical of the higher cross sections associated with hot
fusion reactions compared to cold fusion reactions (Figure \ref{fig1}) for the
synthesis of Z $>$ 112. It is these same cross sections, which challenge our
understanding because current theoretical predictions of the survival
probabilities in these reactions \cite{adamian} would not give cross
sections of this magnitude. For example, Armbruster \cite{arm} using the
best available data on the capture cross sections, the probability of
evolving from the contact configuration to the completely fused system and
the survival probabilities, estimated a evaporation residue production
section for the reaction of 231 MeV $^{48}$Ca + $^{238}$U $\rightarrow
^{286}$112 $\rightarrow ^{283}$112 + 3n of 50 fb.

\section{Experimental}

The reaction $^{238}$U($^{48}$Ca,3n)was studied at the 88-Inch Cyclotron of
the Lawrence Berkeley National Laboratory, using the Berkeley Gas-filled
Separator \cite{victor}. The experimental apparatus was a modified, improved
version of the apparatus used in \cite{victor}, including improved detectors
and data acquisition system, continuous monitoring of the separator gas
purity, and better monitoring of the $^{48}$Ca beam intensity and energy. A 
$^{48}$Ca$^{10+}$beam was accelerated to 243.5 MeV with an average current of 
$\sim$3 x 10$^{12}$ions/s (480 particle-na). The beam went through the 45 
$\mu $g/cm$^{2}$carbon entrance window of the separator before passing
through the $^{238}$U target placed 0.5 cm downstream from the window. The
targets were UF$_{4}$ deposits (U thickness = 463 $\mu $g/cm$^{2}$) with an
0.54 mg/cm$^{2}$Al backing on the upstream side. Nine of the arc-shaped
targets were mounted on a 35-cm wheel that was rotated at 300 rpm. The beam
energy in the target was 228 - 234 MeV \cite{hubert}, encompassing the
projectile energy range used in \cite{yuri4}. The beam intensity was
monitored by two silicon p-i-n detectors (mounted at $\pm $27 degrees with
respect to the incident beam) that detected elastically scattered beam
particles from the target. Attenuating screens were installed in front of
these detectors to reduce the number of particles reaching them (and any
subsequent radiation damage to the detector). The run lasted approximately
5.5 days.

The EVRs (E$\sim$39 MeV) were separated spatially in flight from beam
particles and transfer reaction products by their differing magnetic
rigidities in the gas-filled separator. The separator was filled with helium
gas at a pressure of 96 Pa. The expected magnetic rigidities of 39-MeV 
$^{283}$112 EVRs were estimated using the data of Ghiorso et al.\cite{al}
These estimates were 2.25 Tm from extrapolation of the data in their Figure
\ref{fig4} and 2.63 Tm for their semi-empirical equation (equation 4) for EVR charge.
\ After comparison of the optimum B$\rho $ values determined experimentally
with the BGS for the EVRs from the reaction of 202 MeV $^{48}$Ca with 
$^{176}$Yb, 215 MeV $^{48}$Ca with $^{208}$Pb and 309 MeV $^{64}$Ni with 
$^{208}$Pb which all corresponded to the ``graphical value" of B$\rho $, we
chose a B$\rho $ of 2.25 Tm for the separator magnetic field.

To determine the transport efficiency of the BGS, we used a combination of
measurements and Monte Carlo simulations.  We measured the transport
efficiency of the separator, the efficiency of transporting an EVR produced
in the target and implanting it in the focal plane detector, to be 57\% for
the reaction of 202 MeV $^{48}$Ca with $^{176}$Yb, assuming a cross section
for this reaction of $\sim$790 $\mu $b.  (This latter value was
extrapolated from the measured data of Sahm, et al.\cite{sahm})\ A Monte
Carlo simulation of the separator efficiency for this reaction \cite{ken}
predicted an efficiency of 53\%.  We measured a transport efficiency of 45\%
for the reaction of 215 MeV $^{48}$Ca + $^{208}$Pb $\rightarrow {}^{254}$No
+ 2n.  (This efficiency is based on a cross section for the
$^{208}$Pb($^{48}$Ca, 2n) reaction of 3.0 $\mu $b. \cite{heinz})\ The Monte
Carlo simulation program predicted 51\%.  Having thus ``validated" the Monte
Carlo simulation code, we used it to estimate a transport efficiency for the
$^{283}$112 EVRs of 49\% for the reaction of 231 MeV $^{48}$Ca with 
$^{238}$U under the conditions described above. This value is similar to
efficiencies reported for similar reactions using the Dubna gas-filled
separator. \cite{sub}

As a further demonstration of our ability to measure events similar to those
being sought in the $^{48}$Ca + $^{238}$U experiment, we measured the cross
section for the 215.5 MeV $^{48}$Ca + $^{206}$Pb $\rightarrow {}^{252}$No +
2n reaction by detecting the SF decay (SF branching ratio 0.269) of
$^{252}$No. We measured a cross section of 585 $\pm $90 $\mu $b for this
reaction in agreement with the known value of 500 $\mu $b. \cite{yuri6}

In the focal plane region of the separator, the EVRs passed through a 10 cm
x 10 cm parallel plate avalanche counter (PPAC) \cite{djm} that registered
the time, $\Delta $E, and x,y position of the particles. This PPAC has an
approximate thickness equivalent to 0.6 mg/cm$^{2}$ of carbon. The PPAC was 
$\sim$29 cm from the focal plane detector. The time-of-flight of the EVRs
between the PPAC and the focal plane detector was measured. The PPAC was
used to distinguish between beam-related particles hitting the focal plane
detector and events due to the decay of previously implanted atoms. During
these experiments, the PPAC efficiency for detecting beam-related particles
depositing between 8 and 14 MeV in the focal plane detector was 97.5 -
99.5\%.

After passing through the PPAC, the recoils were implanted in a 32 strip,
300 $\mu $m thick passivated ion implanted silicon detector at the focal
plane that had an active area of 116 mm x 58 mm.  The strips were position
sensitive in the vertical (58 mm) direction. The energy resolution of the
focal plane detector was measured to be $\sim$70 keV (FWHM).  The
differences in measured positions for the $^{252}$No - $^{248}$Fm full
energy $\alpha -\alpha$ correlations in a study of the 215.5 MeV $^{48}$Ca +
$^{206}$Pb reaction had a Gaussian distribution with a FWHM of 0.52 mm 
($\sigma = 0.22$ mm).  The measured position resolution for full energy
alpha particles correlated to ``escape'' alpha particles (which deposited
only 0.5 - 3.0 MeV in the detector) was $\sim$1.2 mm (FWHM).  A second
silicon strip ``punch-through'' detector was installed behind this detector
to reject particles passing through the primary detector. A ``top'' and a
``bottom'' detector were installed in front of the focal plane detector to
detect escaping alpha particles and fission fragments. The focal plane
detector combined with these ``top'' and ``bottom'' detectors had an
estimated efficiency of 75\% for the detection of full energy 10 MeV
$\alpha$-particles following implantation of a $^{283}$112 nucleus.

Any event with E $>$ 0.5 MeV in the focal plane Si-strip detector triggered
the data acquisition. Data were recorded in list mode, and included the time
of the trigger, the position and energy signals from the PPAC and the
Si-strip detectors, and energy signals from the ``top", ``bottom" and
``punch-through" detectors. With the use of buffering ADCs and scalers, the
minimum time between successive events was 15 $\mu $s.

In a study of the 215 MeV $^{48}$Ca + $^{206}$Pb reaction, the pulse height
defect for the $\sim$17 MeV $^{252}$No recoils was determined to be $\sim$10
MeV. This correction was used to determine the expected range of energies
associated with the $\sim$15 MeV $^{283}$112 recoils as they struck the
focal plane detector.

With a beam current of 3 x 10$^{12}$ $^{48}$Ca ions striking the target, the
average total counting rates (E $>$ 0.5 MeV) in the focal plane detector
were $\sim$0.84/s. The average rate of ``alpha particles" (7-14 MeV with no
PPAC signal) was $<$ 0.028/s.

\section{Results and Discussion}

Two search strategies were used to look for events corresponding to the
implantation and decay of $^{283}$112 nuclei.  The first strategy assumed
the decay of $^{283}$112 would occur as outlined in Figure \ref{fig4}, in
accord with the predictions of Smola\'{n}czuk.  We searched for
EVR-$\alpha$, $\alpha$-$\alpha$, and EVR-fission events occurring within 6
s, restricting the range of $\alpha$-particle energies to be from 8 to 11
MeV and the single fragment fission energies to be $\geq$ 90 MeV.  No
events were observed with a total dose of 1.1x10$^{18}$ ions.  This
corresponds to a one-event upper limit cross section of 1.8 pb for
$^{283}$112 nuclei decaying by alpha-particle emission and 1.6 pb for
spontaneously fissioning $^{283}$112 nuclei when one takes into account the
differing efficiencies of detecting fission fragments and alpha-particle
decay chains.

A second strategy involved searching for events similar to those reported by
the Dubna-GSI-RIKEN group. \cite{yuri4}. We searched for EVR-$\alpha$, 
$\alpha$-$\alpha$, and EVR-fission events occuring within 1000 s, using the
same energy restrictions as in the first search.  No EVR-fission events were
found, leading to a one event upper limit cross section of 1.6 pb for the
type of event reported by the Dubna-GSI-RIKEN group or any chain terminating
in an SF decay.  Due to a significant number of accidental EVR-$\alpha$ and
$\alpha$-$\alpha$ events, no meaningful upper limit could be set for
EVR-$\alpha$ events with these longer correlation times.

The one event upper limit cross section for the production of spontaneously
fissioning $^{283}$112 nuclei of 1.6 pb is just below that reported by the
Dubna-GSI-RIKEN group of 5.0$_{-3.2}^{+6.3}$ pb. Another relevant
observation is that of Yakushev, {\it et al.}, \cite{yaku} who reported the
failure to observe any spontaneously fissioning $^{283}$112 nuclei in the
reaction of 234 MeV $^{48}$Ca with $^{238}$U using the assumption that
element 112 behaves like Hg, a volatile liquid, in its chemistry. If element
112 behaves chemically like Hg, then this observation would suggest an upper
limit cross section of $\sim$1.5 pb for this reaction. An alternative
explanation \cite{yaku,pitzer} is that element 112 behaves chemically like a
noble gas (Rn). Recent theoretical predictions \cite{antonenko} using the
dinuclear system approach, have suggested a cross section for the 
$^{238}$U($^{48}$Ca,3n)$^{283}$112 reaction of 1.7 pb.

Further work is needed to establish the cross section for the production of 
$^{283}$112 in the $^{238}$U($^{48}$Ca,3n) reaction. Because the reported
spontaneous fission decay is not definitive to determine the Z and A of this
nucleus, it seems especially important to detect the $\alpha$-decay branch
for this nuclide. The apparently small cross sections and/or weaker
$\alpha$-decay branching ratios make this worthwhile effort difficult. If,
as indicated in this work, the production cross section for $^{283}$112 in
the $^{238}$U($^{48}$Ca,3n) reaction is $\sim$2 pb or less, then it becomes
more difficult to understand the reported cross sections of $\sim$1 pb for
the production of elements 114 and 116 in similar reactions.

We gratefully acknowledge the operations staff of the 88-Inch Cyclotron and
its ion source person, Daniela Leitner, for providing intense, steady beams
of $^{48}$Ca. We thank Victor Ninov, Z. Huang, and T.N. Ginter for their
invaluable assistance during this experiment. Financial support was provided
by the Office of High Energy and Nuclear Physics, Nuclear Physics Division
of the U.S. Dept. of Energy, under contract DE-AC03-76SF00098 and grant
DE-FG06-97ER41026 and the Swedish Research Council.

\end{document}